\documentclass[%
superscriptaddress,
longbibliography,
amsmath,amssymb,
prf,]{revtex4-1}
\usepackage[export]{adjustbox}
\usepackage{mathtools}
\usepackage{graphicx}% Include figure filessn
\usepackage{lineno}
\usepackage{color}% Include figure files
\usepackage{xcolor}% Include figure files
\usepackage{wasysym}% Include figure files
\usepackage{dcolumn}% Align table columns on decimal point
\usepackage{bm}% bold math
\usepackage[colorlinks = true,
            linkcolor = blue,
            urlcolor  = blue,
            citecolor = blue,
            anchorcolor = blue]{hyperref}\usepackage{subfigure}
\usepackage{comment}

\begin{document}
\title{Non-intrusive, transferable model for coupled turbulent channel-porous media flow based upon neural networks}

\author{Xu Chu}
%\email{xu.chu@simtech.uni-stuttgart.de}
\affiliation{Cluster of Excellence SimTech, University of Stuttgart, Germany}

%\author{J\"org Schumacher}
%\affiliation{Institute of Thermodynamics and Fluid Mechanics, Technische Universit{\"a}t Ilmenau, D-98684 Ilmenau, Germany}
%\affiliation{Tandon School of Engineering, New York University, New York, NY 11201, USA.}

\author{Sandeep Pandey}
\email{sandeep.pandey@tu-ilmenau.de}
\affiliation{Institute of Thermodynamics and Fluid Mechanics, Technische Universit{\"a}t Ilmenau, D-98684 Ilmenau, Germany}

\date{\today}

\begin{abstract}

Turbulent flow over permeable interface is omnipresent featuring complex flow topology. In this work, a data-driven, end-to-end machine learning model has been developed to model the turbulent flow in porous media. For the same, we have derived a non-linear reduced order model with a deep convolution autoencoder network. This model can reduce highly resolved spatial dimensions, which is a prerequisite for direct numerical simulation, by 99\%. A downstream recurrent neural network has been trained to capture the temporal trend of reduced modes, thus it is able to provide future evolution of modes. We further evaluate the trained model's capability on a newer dataset with a different porosity. In such cases, fine tuning could reduce the efforts (up to two-order of magnitude) to train a model with limited dataset (10\%) and knowledge and still show a good agreement on the mean velocity profile. Especially, the fine-tuned model shows a better agreement in the porous domain than the channel and interface areas indicating the topological feature is less challenging for training than the multi-scale nature of the turbulent flows. 
Leveraging the current model, we find that even quick fine-tuning—achieving an impressive order-of-magnitude reduction in training time by approximately $\mathcal{O}$(10$^2$)—still results in effective flow predictions. This promising discovery encourages the fast development of a substantial amount of data-driven models tailored for various types of porous media. The diminished training time substantially lowers the computational cost when dealing with changing porous topologies, making it feasible to systematically explore interface engineering with different types of porous media.
Overall, the data-driven model shows a good agreement, especially for the porous media which can aid the DNS and reduce the burden to resolve this complex domain during the simulations. 
The fine tuning is able to reduce the training cost significantly and maintain an acceptable accuracy when a new flow condition comes into play.
%Based on the existing mode, even a rapid fine-tuning with a $\mathcal{O}$(10$^2$) reduction of the training time is able to deliver an efficient prediction of the flow. This exciting finding motives to build up a series of data-driven models for cases with different porous media since the training cost for alternating porous topology is significantly reduced. It can serve for a systematic investigation of interface engineering with porous media.

\end{abstract}

\maketitle

%**************************************************************
\section{Introduction}
\label{sec:1}

Turbulent flow across permeable interfaces is a phenomenon that is pervasive in both natural environments and engineering contexts, manifesting in diverse applications ranging from sediment transport to transpiration cooling in gas turbines. Porous media are often characterized by a complex spatial topology \citep{Chu.2018a, lan2021all,Chu.2019}, and numerous properties have been recognized to exert a substantial influence on both mass and momentum transport at the interface between porous and free-flowing media. Despite this understanding, the task of comprehending and optimizing the macroscopic and microscopic characteristics of porous media continues to present a challenge. The underlying reasons for this complexity include the inherent difficulties associated with conducting experimental measurements at the pore scale \cite{Terzis.2019} and the substantial computational resources required for numerical simulations that adequately resolve the entire range of relevant scales.
Historical examinations of turbulent flow over permeable substrates have investigated the impact of variable permeability on surface flows. In scenarios where interfacial permeability is low, the turbulent surface flow exhibits characteristics akin to a canonical boundary layer, owing to the limited disruption of near-wall structures. Conversely, when permeability is augmented, large-scale vortical structures begin to manifest within the surface flow. This phenomenon has been ascribed to Kelvin-Helmholtz (KH) type instabilities, originating from the inflection points present in the mean velocity profile. In more recent times, the exploration of porous media as a means of achieving drag reduction has catalyzed a series of comprehensive studies focused on the subject of anisotropic permeability.

Direct Numerical Simulation (DNS) provides a distinct advantage for observing and analyzing the physics of turbulence within a constrained small spatial domain, extending beyond traditional applications such as channel \citep{Pandey.2020} and pipe flows \cite{Chu.2016, Pandey.2017b}. \citet{Suga.2020} conducted an investigation into the influence of the anisotropic permeability tensor within porous media at a regime characterized by higher permeability. Their findings revealed that both streamwise and spanwise permeabilities contribute to turbulence enhancement, while vertical permeability alone does not exert a similar influence. The effect on turbulence is particularly pronounced in the presence of porous walls exhibiting streamwise permeability, as this configuration promotes the development of large-scale streamwise perturbations induced by Kelvin–Helmholtz instability.
However, despite the insights offered by DNS, its utilization for extensive parametric studies, often required in industrial applications, remains a costly solution. In such contexts, it is still customary to employ either analytical models or advanced turbulence models, which tend to provide satisfactory accuracy within established applications. Nevertheless, these models are not without their challenges, as they can exhibit divergent behavior, and the calibration process can prove to be both intricate and time-consuming.

Machine learning (ML) is on surge where it has successfully tackled many complex tasks ranging from self-driving cars \cite{jing2020self}, drug discovery \cite{vamathevan2019drugs}, weather prediction \citep{chattopadhyay2020predicting}, and more recently state-of-the-art large language model such as GPT for natural language processing \citep{zhang2021commentary}. Machine learning has been democratized heavily in the last decade and some of the reasons include the vast availability of open-source libraries, an active community, a huge volume of incoming data from physical as well as numerical experiments, and an ease of access to high-performance computing resources via cloud. In recent years, ML has shown potential in the modeling of various fluid flow use-cases \citep{pandey2022direct, heyder2022generalizability, nakamura2021convolutional, fang2019neural, sayyari2022unsupervised, chu2018computationally, larranaga2023data,pandey2020perspective,pfeffer2022hybrid,heyder2021echo,Chang.2018}. Typically, big data ML problems rely upon some kind of feature extraction process e.g. conventional feature selection process, long-standing proper orthogonal decomposition (POD) \cite{Wang.2022,liu2023interfacial,liu2023large}, relatively new dynamic mode decomposition (DMD) \cite{kutz2016dynamic} or ML-based methods \citep{storcheus2015survey}. Feature extractors transform the input raw data into information-preserving feature vectors with the primary goal of reducing the dimensionality of the data. Several attempts have been made in the past few years to combine POD or DMD with machine learning \citep{pandey2020reservoir, heyder2021echo, ghazijahani2022benefits, jia2022hybrid}. However, these methods have certain limitations especially when the flow advances in the turbulent flow regimes. Therefore, an end-to-end machine learning system is being suggested where POD and DMD are replaced by a non-linear DNN \citep{pandey2020reservoir, nakamura2021convolutional, rahman2019nonintrusive}. A common architecture is a deep convolution autoencoder (DAE) to extract the features for further processing and decision making \citep{ryu2019convolutional, polic2019convolutional}. Turbulent flow problem often involves the tempo-spatial data, therefore, extracted features could be used to train a downstream network to build a data-driven dynamic model. A common choice is recurrent neural network (RNN) which preserves the sequential relationship in a time-series \cite{srinivasan2019predictions}. A combination of such approaches has shown exemplary results in terms of data compression and modeling \cite{nakamura2021convolutional}.

In this work, we make an attempt to model complex flow behaviors in a porous medium with the help of a data-driven approach. The flow is in a turbulent regime and due to the presence of porosity, geometry is resolved to a finer scale. This setup makes the end-to-end ML task challenging and equally interesting. The goal is to combine the deep learning-based feature extraction step followed by a recurrent neural network to model temporal evolution. Echo state network was chosen as the RNN due to its proven efficacy in modeling multi-dimensional tempo-spatial data. 
The trained models are restricted to the domain of training and work poorly beyond the trained data domain which could be a specific parameter such as the Reynolds number in fluid flow problems. A commonly used strategy called transfer learning is typically used, where the objective is to transfer the gained knowledge from the training of source task to related target task with a limited dataset \cite{zhuang2020comprehensive}. 
Fine-tuning, which can be considered a part of transfer learning, is one methodology where an already trained and optimized model is tuned with another similar set of data for a similar task. This could decrease the training time significantly in addition to neural architecture search and training data requirements \cite{tajbakhsh2016convolutional}. Fine-tuning can be performed on the entire network level where we are allowed to retrain all the layers while initializing the parameters with the pre-trained model. Alternatively, several layers can be frozen and retraining is allowed on only a few layers. The former method is suitable when the network size is very large and data is limited. While the first approach could lead to overfitting \cite{yosinski2014transferable}. Therefore, we also present the result from fine-tuning where a trained model from a given data domain was transferred to a different data domain.  

The remainder of the paper is divided into 4 sections. Section \ref{synthetic_data_generation} describes the fluid flow problem in porous media along with the underlying governing equations and numerical method. Section \ref{ml_framework} explains in detail regarding the hierarchical data-driven model. Section \ref{results} describes the results from training and fine-tuning. With section \ref{concl}, we conclude the work and give a brief outlook.

%**************************************************************
\section{Synthetic data generation}
\label{synthetic_data_generation}

%In our DNS, the three-dimensional incompressible Navier$-$Stokes equations are solved in non-dimensional form, 

%\begin{equation}
%\frac{\partial u_j}{\partial x_j}=0
%\label{eq:1}
%\end{equation}
%\begin{equation}
%\frac{\partial u_i}{\partial t}+\frac{\partial u_i u_j}{\partial x_j}=-\frac{\partial p}{\partial x_i}+\frac{1}{Re_D}\frac{\partial^2 u_j} {\partial x_i \partial x_j}+\Pi \delta_{i1}
%\label{eq:2}
%\end{equation}

%where $\Pi$ is a constant pressure gradient in the mean-flow direction.  The governing equations are normalized using the half-width of the whole simulation domain $H$(figure \ref{fig:setup}a) and the averaged bulk velocity $U_b$ of the channel region ($y/H=[0,1]$). Hereafter, the velocity components in the streamwise $x$, wall-normal $y$, and spanwise $z$ directions are denoted as $u$, $v$, and $w$, respectively.  The domain size ($L_x/H \times L_y/H \times L_z/H$) is $10 \times 2 \times 0.8\pi$ in all cases. The lower half ($y/H=[-1,0]$) contains the porous media, and the upper half ($y/H=[0,1]$) is the free channel flow. The   porous layer consists of 50 cylinder elements along the   streamwise direction and 5 rows in the wall-normal   direction, as illustrated in figure \ref{fig:setup}. The distance $D$ between two nearby cylinders is fixed at $D/H=0.2$. No-slip boundary condition is applied to the cylinders, the upper wall, and the lower wall. Periodic boundary conditions are used in both streamwise and spanwise directions. 

High-fidelity simulations such as Direct Numerical Simulation (DNS) \citep{Chu.2020, Pandey.2020, Chu.2021, Chu.2021b} can serve reliable spatial and temporal resolved data for the modeling. In our DNS, the three-dimensional incompressible Navier$-$Stokes equations are solved in a non-dimensional form as follows:

\begin{equation}
\frac{\partial u_j}{\partial x_j}=0
\label{eq:1}
\end{equation}
\begin{equation}
\frac{\partial u_i}{\partial t}+\frac{\partial u_i u_j}{\partial x_j}=-\frac{\partial p}{\partial x_i}+\frac{1}{Re_D}\frac{\partial^2 u_j} {\partial x_i \partial x_j}+\Pi \delta_{i1}
\label{eq:2}
\end{equation}

\begin{figure}[b!]
	\begin{tabular}{c}
	 \includegraphics[width=0.7\textwidth]{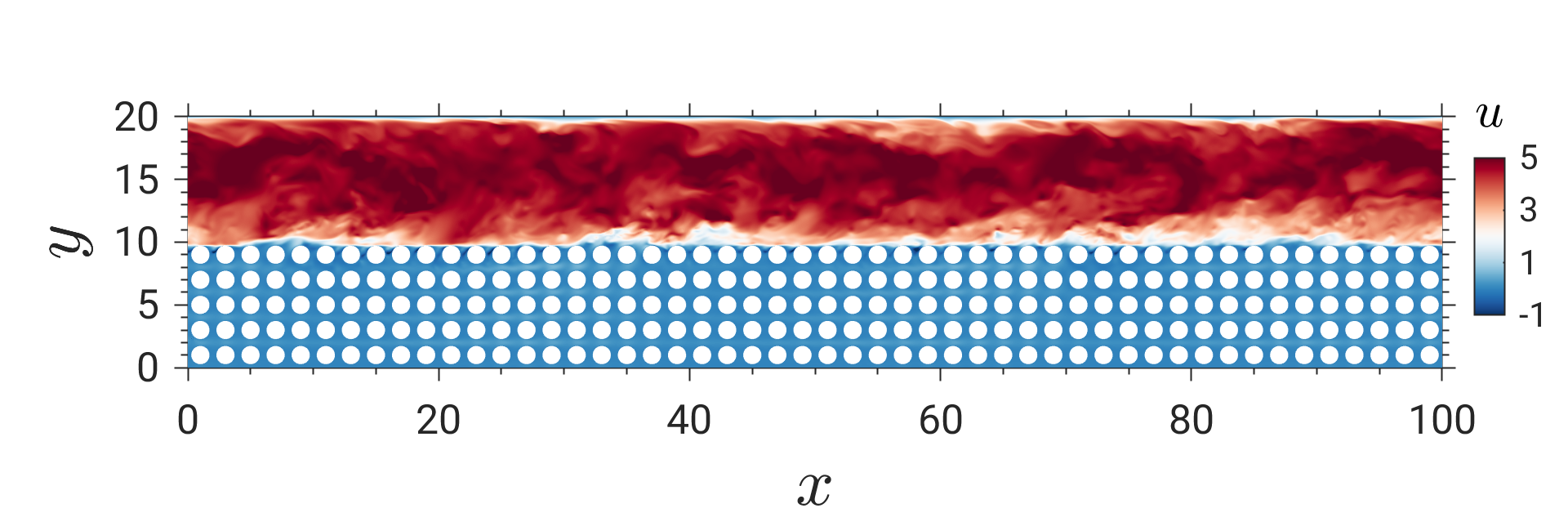}    \\
	\end{tabular}
	\caption{A $x-y$ cutting plane of the computational domain with intastanteous streamwise velocity is depicted in color.}
	\label{fig:setup}
\end{figure}

Here, $\Pi$ represents a constant pressure gradient in the mean-flow direction. The governing equations are normalized by employing the half-width of the entire simulation domain, denoted as $H$ (refer to figure \ref{fig:setup}a), and the mean bulk velocity $U_b$ within the channel region specified by $y/H=[0,1]$. The velocity components in the streamwise ($x$), wall-normal ($y$), and spanwise ($z$) directions are henceforth expressed as $u$, $v$, and $w$, respectively. The domain dimensions ($L_x/H \times L_y/H \times L_z/H$) are fixed at $10 \times 2 \times 0.8\pi$ for all scenarios considered. The lower half of the domain ($y/H=[-1,0]$) encompasses the porous media, while the upper half ($y/H=[0,1]$) constitutes the free channel flow. The porous layer is constructed with 50 cylindrical elements aligned in the streamwise direction and 5 rows positioned in the wall-normal direction, as delineated in figure \ref{fig:setup}. The distance $D$ between two adjacent cylinders is consistently set at $D/H=0.2$. No-slip boundary conditions are enforced on the cylinders, the upper wall, and the lower wall, whereas periodic boundary conditions are implemented in both the streamwise and spanwise directions.

%The simulations are first run for a total of 10 flow through times for developing the flow and then another for 5 flow through times to collect statistics \textcolor{red}{(?? is this long enough?)}.

%The geometry is discretized using quadrilateral elements on the $x$-$y$ plane with local refinement near the interface. Modified Lagrange polynomials are used within each element on the $x$-$y$ plane. 

%The spectral/$hp$ element solver Nektar++ is used to solve equations (\ref{eq:1}, \ref{eq:2}) \citep{Cantwell.2011, Chu.2019, Chu.2020,Pandey.2020, Chu.2021, Chu.2021b, Wang2021assessment}. The geometry in the $x$-$y$ plane is discretized into quadrilateral elements with local refinement near the cylinders (see figure \ref{fig:setup}a). Local element expansions are applied based on the modified Legendre basis \citep{karniadakis2005spectral}. We used flexible polynomial orders across the wall normal range in a continuous Galerkin projection. The polynomial order in the free flow region $y/H=[0.2, 1]$ is $P=6$-$7$. The near wall region and the top two layers of cylinders $y/H=[-0.4, 0.2]$ are enhanced with a higher order of $P=8$-$9$. A lower order of $P=5$ is selected in the deeper positions of the cylinder array ($y/H=[-1, -0.4]$). The spanwise direction is extended with a Fourier spectral method. The 2/3 rule is used to avoid aliasing errors. The time-stepping is performed with a second-order mixed implicit-explicit (IMEX) scheme proposed by \cite{karniadakis1991high}. The time step is fixed at $\Delta T/(H/U_b)=5\times10^{-4}$.

The spectral/$hp$ element solver Nektar++ is utilized to solve the equations as referred to in Eqn. \ref{eq:1}-\ref{eq:2} \citep{Cantwell.2011, Chu.2019, Wang2021assessment,Wang2021}. The geometric configuration in the $x$-$y$ plane is discretized utilizing quadrilateral elements, with localized refinement in the vicinity of the cylinders (refer to figure \ref{fig:setup}(a)). Localized element expansions are implemented based on the modified Legendre basis \citep{karniadakis2005spectral}. Flexibility in polynomial orders is employed across the wall-normal range through continuous Galerkin projection. Specifically, the polynomial order within the free-flow region, defined as $y/H=[0.2, 1]$, is set at $P=6$-$7$. The proximal wall region and the upper two strata of cylinders, where $y/H=[-0.4, 0.2]$, are augmented with a superior order of $P=8$-$9$. Conversely, an inferior order of $P=5$ is designated within the more profound locations of the cylinder array ($y/H=[-1, -0.4]$). The spanwise orientation is broadened utilizing a Fourier spectral method, and the 2/3 rule is implemented to circumvent aliasing errors. The temporal progression is executed with a second-order mixed implicit-explicit (IMEX) scheme, with the time step fixed at $\Delta T/(H/U_b)=5\times10^{-4}$.

\begin{table}
\centering
\begin{tabular}{ccccccccccc}

Case&$\varphi$  & 
$Re_{\tau}^{p}$&$Re_{\tau}^{s}$&$\sqrt{K_{xx}}^{p+}$,$\sqrt{K_{yy}}^{p+}$ 
 &$\sqrt{K_{zz}}^{p+}$ & $C_{xx}^{{p+}},C_{yy}^{{p+}}$ & $r_c^{{p+}}$& $C_f^p$ & $C_f^s$\\

C05  &0.5 & 336  & 180  &4.55&8.86&0.37&42&0.0112&0.0084\\
C06  & 0.6 & 464   & 190  &9.34&15.23&0.58&48&0.0149 &0.0085 \\
%C07   &0.7   & 625   & 160  &20.65&30.98&2.99&52 &0.0278 & 0.0096\\
%C08   &0.8   & 793   & 170  &36.83&53.99&13.58&52 & 0.0298 & 0.0094\\

\end{tabular}
\caption{Simulation parameters. The porosity of the porous medium
  region is $\varphi$. The friction Reynolds numbers are $Re_{\tau}^{p}$ and
  $Re_{\tau}^{s}$ for the porous and impermeable top walls respectively. $\sqrt{K_{\alpha\alpha}}^{p+}$ and $C_{\alpha\alpha}^{{p+}}$ are the diagonal components of the permeability tensor and Forchheimer coefficient, respectively, in the direction of $\alpha$ ($\alpha\in\{x, y,z \}$), which are normalized by wall units. $r_c^{{p+}}$ is the radius of the cylinders.}
\label{tab:1}
\end{table}

Two DNS cases are performed with varying porosity $\varphi=0.5,0.6,0.7$, and $0.8$, which is defined as the ratio of the void volume to the total volume of the porous structure. The parameters of the simulated cases are listed in TABLE \ref{tab:1}, where the cases are named after their respective porosity. The superscripts $(\cdot)^p$ and $(\cdot)^{s}$ represent permeable wall and smooth wall side variables, respectively. Variables with superscript $^+$ are scaled by friction velocities $u_\tau$ of their respective side and viscosity $\nu$.

It should be noted that the distance between the cylinders remains constant, while the porosity undergoes modification through variation in the cylinder radii. The normalized cylinder radius is found to be within the range $r_c^{p+}=r_c u_\tau^{p}/\nu=42$-$52$ for all examined cases (refer to TABLE \ref{tab:1}), thereby implying that the effects of surface roughness are presumed to be consistent across different scenarios. For all instances, the Reynolds number associated with the top wall boundary layer is configured at $Re_\tau^{s}=\delta^{s}u_\tau^{s}/\nu\approx180$ ($\delta$ represents the distance between the position of maximal streamwise velocity and the wall). This configuration aids in minimizing fluctuations in the top wall boundary layer. In the region of the upper smooth wall, the streamwise cell size spans from $4.1\le\Delta x^{s+}\le6.3$, and the spanwise cell size is confined to below $\Delta z^{s+}=5.4$. On the side corresponding to the porous media, the value of $\Delta z^{p+}$ is capped at 8.4, whereas $\Delta x^{p+}$ and $\Delta y^{p+}$ are augmented by polynomial refinement of the local mesh \citep{Cantwell.2011}. The cumulative number of grid points fluctuates between 88$\times10^6$ (C05) and 110$\times10^6$ (C06), with each cylinder within the porous domain resolved using 80 to 120 grids along its perimeter.
The spatial resolution of the current work aligns closely with that of preceding DNS studies \cite{Shen.2020,Karra.2022}. Additionally, the utilization of a high-order scheme in conjunction with body-fitted mesh confers a notable advantage in resolving fine scales and necessitates fewer grids compared to both finite volume and immersed boundary methods to attain equivalent accuracy \citep{Kooij.2018}. Furthermore, a comparative analysis of our grid resolutions with the Kolmogorov length scale, denoted as $\eta=(\nu/\epsilon)^{1/4}$, at the interface has been conducted. For all cases, the findings show that $(\Delta x/\eta){y=0}\leqslant2$, $(\Delta y/\eta){y=0}\leqslant1$, $(\Delta z/\eta)_{y=0}\leqslant4.5$, thereby validating that the current resolution is indeed sufficient

%**************************************************************
\section{Data-driven modelling}
\label{ml_framework}

\begin{figure}[ht!]
\centering
\includegraphics[width=17cm, keepaspectratio]{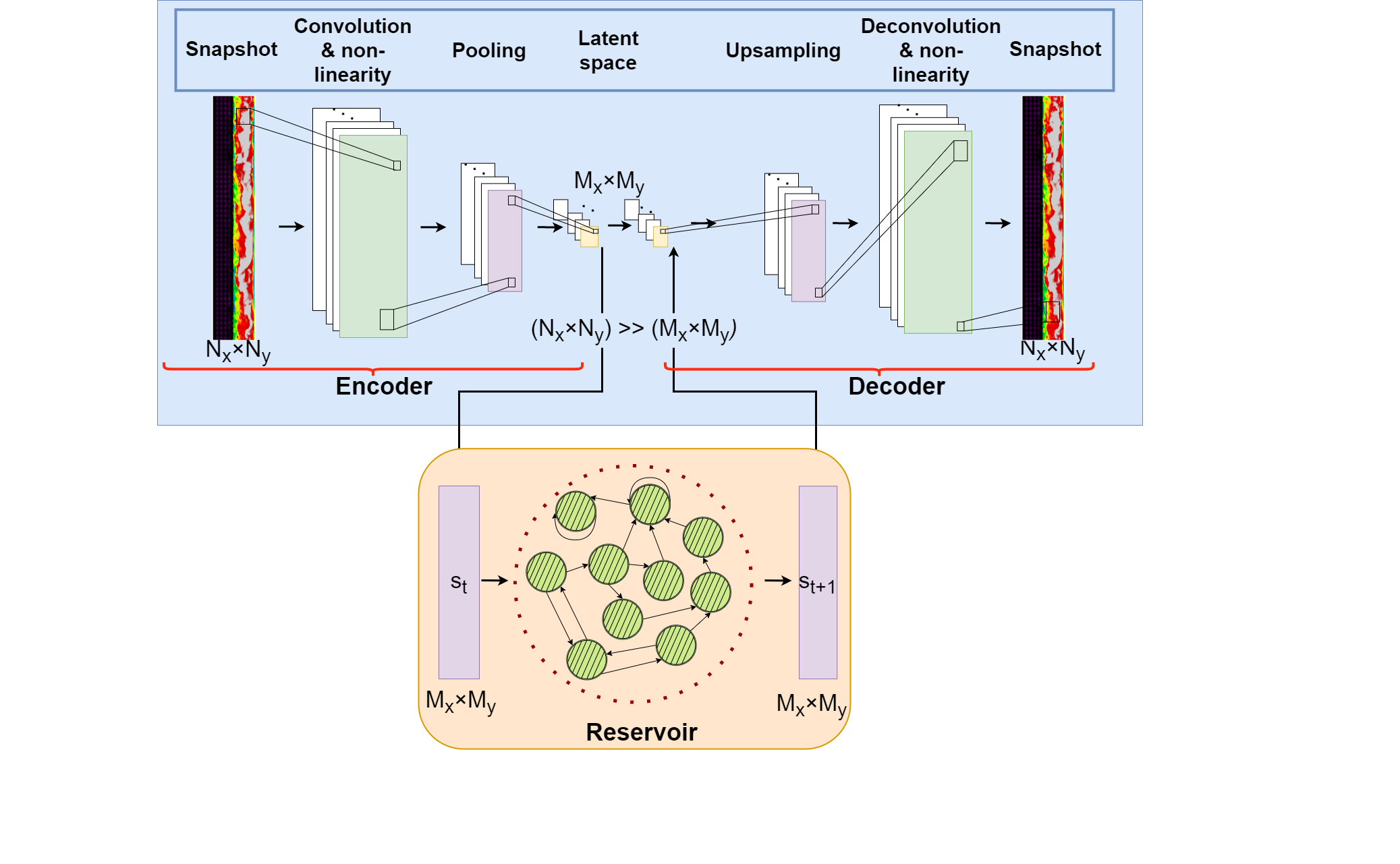}
\caption{\label{auto_results} Two-step machine learning architecture for end-to-end ROM model for porous media. In the first step, DAE is trained as a feature extractor cum decoder while the second step trains the extracted feature in an autoregressive manner.}
\label{Fig-2}
\end{figure}

Data driven modelling (DDM) is an approach to harness the available data about
a system and establish a connection between the system state variables (input,
internal and output variables) without explicit knowledge of the physical behaviour \cite{solomatine2008data}. In this work, we have the synthetic data derived from the DNS experiments while the end model should be based upon this data and should able to provide a model to predict the temporal evolution of a state variable. Figure \ref{Fig-2} depicts our end-to-end 2-step ML framework. To have an extendable framework, we chose to first extract the features with the help of a deep encoder and feed the extracted features to a second-level recurrent neural network (RNN). This architecture is inspired from our earlier work \cite{pandey2022direct}. 

%------------------------------------------------------
%------------------------------------------------------

\subsection{Feature extraction with autoencoder}
\label{dae}

%The deep convolution autoencoder (DAE) is an implementation of an unsupervised machine learning paradigm, where the training of the model does not rely upon the labeled data.  The objective of learning is to extract efficient representations or features in a low-dimension hyper-plane that can reproduce the data with minimal reconstruction error.  That is to say, an autoencoder encodes input data $\bm{I}$ into $\bm{c}=\textnormal{encode}(\bm{I})$ with $\bm{c} \in \mathbb{R}^l$. And it decodes $\bm{c}$ back into $\hat{\bm{I}}=\textnormal{decode}(\bm{c})$ subsequently. The reconstruction error is calculated with $\hat{\bm{I}}-\bm{I}$. Here, 2-dimensional data were obtained from the DNS as described in \ref{synthetic_data_generation}. Thus autoencoder consists of several convolution layers resulting in a deep architecture. Each convolution layer in the network acts as a local feature extractor and shares the weights. The convolution layer is followed by a pooling layer, max pooling in this case, which reduces the dimension and focuses on the non-repeating key features. It gives a unique property of translation invariance to the network. At the end of the encoder network, there is a bottleneck layer that has the latent space or extracted features. These features are then passed to a decoder section which will attempt to recreate the original DNS data from the features. The training starts with random initialization of all the weights in the network and learns them as per the loss function via backpropagation. 

The deep convolution autoencoder (DAE) operates within the framework of unsupervised machine learning, meaning that the model's training does not depend on labeled data. The learning objective is to distill efficient representations or features within a low-dimensional hyper-plane, with the aim of reproducing the original data while minimizing reconstruction error.
In more formal terms, an autoencoder encodes input data $\bm{I}$ into a compact form $\bm{c}=\textnormal{encode}(\bm{I})$, where $\bm{c} \in \mathbb{R}^l$. It then decodes this compact form back into a reconstructed representation $\hat{\bm{I}}=\textnormal{decode}(\bm{c})$. The difference between the reconstructed and original data, given by $\hat{\bm{I}}-\bm{I}$, constitutes the reconstruction error.

In the specific implementation discussed, 2-dimensional data were sourced from DNS, as delineated in \ref{synthetic_data_generation}. The architecture of the autoencoder is deep, comprising several convolution layers. Each layer functions as a local feature extractor and shares its weights across the network. Following each convolution layer is a pooling layer—specifically, a max pooling layer in this instance—which diminishes the dimensionality of the data and emphasizes non-repetitive key features. This contributes to the network's unique translation invariance property.
The encoder segment of the network concludes with a bottleneck layer, which harbors the latent space or extracted features. These features are subsequently channeled into a decoder section, which attempts to reconstruct the original DNS data. The training process initiates with random weight initialization across the network and iteratively refines these weights according to the loss function, employing backpropagation for optimization.

Consider three-dimensional input data $\bm{I} \in \mathbb{R}^{ N_x \times N_y \times C_\textrm{in}}$ with $C_{\textrm{in}}$ because the number of input channels. Here, $C_{\textrm{in}}=1$ as we load the streamwise velocity field $u_x$ only into the network and thus $\bm{I} \in \mathbb{R}^{ N_x \times N_y }$.  This input is convoluted with a kernel $k_m$ as shown in Eqn. \ref{DAE}.

\begin{equation}
    \textrm{Conv}(m,\bm{I})=\psi \left( b_m+ \sum_{i=1}^{C_\textrm{in}}k_m \ast \bm{I}_i \right) \,\,\, \textrm{for} \, m \in M
    \label{DAE}  
\end{equation}

In the encoder, multiple convolutional layers are typically followed by pooling layers.  A window of configurable size and sliding is used with a configurable step across the input. This process combines the input within each window to produce a single output element using a chosen aggregation function. The common max-pooling layer's aggregation function retains only the maximum element within each window. The pooling layer's step size is usually chosen to reduce the input's dimensionality, denoted as $\bm{I}$.

%As a result of the encoding process, the eventual output of the encoder resides in the latent space $\mathbb{R}^l$, where $l \le N_x \times N_y$, with $N_x$ and $N_y$ being the dimensions of the input. The decoder of the convolutional autoencoder (CAE) follows a mostly analogous design, with one main difference: it employs upsampling layers instead of max-pooling layers to decode the data back to its original input domain size. The upsampling step, also known as unpooling, doubles the data dimension. In the present work, nearest-neighbor linear interpolation is used for the upsampling process. Effectively processing complex turbulence data necessitates a deep architecture consisting of multiple convolutional and accompanying upsampling layers. This architecture ensures obtaining a representation without substantial information loss. In the end-to-end scenario, after training all networks, we employ the decoder of the DAE to derive a compressed form $\hat{\bm{c}}_t$ from a given turbulence snapshot $\bm{I}_t$. This compressed snapshot then becomes the input to a subsequent Recurrent Neural Network (RNN), which forecasts the next reduced representation $\hat{\bm{c}}_{t+1}$. The decoder unit of the DAE decodes $\hat{\bm{c}}_{t+1}$ back into the input domain, reconstructing the forecasted turbulence snapshot. By utilizing deep convolutional autoencoders integrated with recurrent neural networks, we achieve efficient and informative compression of complex turbulence data, enabling more effective forecasting and analysis in the field of fluid mechanics.

Within the encoder segment of a DAE, multiple convolutional layers are commonly succeeded by pooling layers. These layers operate by deploying a configurable sliding window across the input, aggregating the information within each window into a single output element via a specified aggregation function. A frequent choice for this function is the retention of the maximum element within each window, as implemented in a max-pooling layer. This approach serves to diminish the dimensionality of the input, represented by $\bm{I}$, typically through the strategic selection of the pooling layer's step size.

The cumulative effect of the encoding sequence is to map the input data into a latent space $\mathbb{R}^l$, where $l \le N_x \times N_y$, and $N_x$ and $N_y$ denote the dimensions of the initial input.
Following the encoder, the decoder of the DAE primarily mirrors the encoder's design, with a critical distinction: it substitutes max-pooling layers with upsampling layers to expand the data back to its original domain size. This expansion process, known as unpooling, essentially doubles the data's dimensions. In the context of the work discussed, nearest-neighbor linear interpolation is employed to facilitate the upsampling.
The processing of intricate turbulence data necessitates a sophisticated, deep architecture composed of multiple convolutional layers, augmented by corresponding upsampling layers. Such an arrangement ensures the attainment of a faithful representation, minimizing information loss.
In the comprehensive system, subsequent to training all relevant networks, the DAE's decoder is applied to extract a compressed form, denoted by $\hat{\bm{c}}_t$, from a specific turbulence snapshot $\bm{I}_t$. This condensed snapshot is then input into an RNN, which predicts the subsequent reduced representation $\hat{\bm{c}}_{t+1}$. The DAE's decoder translates $\hat{\bm{c}}_{t+1}$ back into the input domain, recreating the anticipated turbulence snapshot.
Through the integrated use of deep convolutional autoencoders and recurrent neural networks, the system attains efficient and insightful compression of complex turbulence data. This approach enhances the capacity for predictive modeling and nuanced analysis within the realm of fluid mechanics.

%In the encoder, multiple convolutional layers are typically followed by pooling layers. A window of configurable size and sliding is used with a configurable step across the input. Thereby, the input per window is aggregated into a single output element with a selectable aggregation function. This aggregation function of the common max-pooling layer retains only the maximum element per window. The pooling layer’s step size is typically chosen such that it reduces the dimensionality of the input $\bm{I}$. The eventual output of the encoder consists of data in the latent space $\mathbb{R}^l$ where $l \le N_x \times N_y $. The CAE’s decoder follows a mostly analogous design. The main difference is the use of upsampling instead of max-pooling layers to decode the data back to input domain size. An upsampling step, also known as unpooling, doubles the data dimension. Nearest-neighbour linear interpolation is therefore used in the present work. The effective processing of complex turbulence data requires a deep architecture consisting of multiple convolutional and accompanying upsampling layers in order to obtain a representation without substantial loss of information. In the end-to-end scenario and after training all networks, we utilize the DAE’s encoder to derive a compressed form $\hat{\bm{c}}_t$ from a given snapshot  $\bm{I}_t$. The compressed snapshot becomes input to a subsequent RNN that forecasts the next reduced representation $\hat{\bm{c}}_{t+1}$, which is then decoded back into the input domain by the decoder unit of the DAE.

\subsection{Echo State Network (ESN)}
\label{esn}

Within the framework of the present study, ESN serves as a downstream recurrent neural network, functioning to predict the temporal evolution of the features extracted by the encoder segment of a DAE in an autoregressive fashion. The forecasted values are reintroduced to the decoder segment of the DAE, facilitating the recreation of the target feature. During the operational phase, the focus is exclusively on the trained ESN and the decoder, while the encoder is disregarded.

The structural construct of the reservoir is represented as an $N \times N$ adjacency matrix, denoted by $\textbf{W}_{\eta}^{(r)}$. The initialization of this reservoir matrix is predicated on a specific vector of hyperparameters, symbolized by $\eta$, and is applied to translate the ESN's input into a concealed representation known as the reservoir state, amassing information from previous inputs. The reservoir dimension, referred to as $N$, is usually selected to substantially exceed the latent space's dimension, following the criterion $N \ge l$. Additionally, this reservoir is dynamically updated with every time increment, receiving new input. Specifically, at each discrete time $t$, the input vector $\bm{c}_t$ modulates the computation of an updated reservoir state $\bm{r}_t \in \mathbb{R}^N$.

The evolution of the reservoir state is defined by:
\begin{equation}
\bm{r}t=(1-\alpha)\bm{r}{t-1}+\alpha \textrm{tanh} \left(\textbf{W}^{(\textrm{in})}\bm{c}t +\textbf{W}^{(\textrm{r})}\eta\bm{r}{t-1} \right)
\end{equation}
where $\alpha$ denotes a leakage rate, dictating the integration of previous states and current inputs. The random matrices $\textbf{W}^{(\textrm{in})}$ and $\textbf{W}^{(\textrm{r})}\eta$ are initialized at the commencement of training and are retained as constants. The ESN's output at the time step $t$ is deduced by:
\begin{equation}
\hat{\bm{c}}_{t+1}= \textbf{W}^{(\textrm{out})} \bm{r}_t
\end{equation}

Contrasting with convolutional networks, in the ESN case, only the final output layer requires training, obviating the need for backpropagation across several epochs. This implies that the elements of the output matrix $\textbf{W}^{(\textrm{out})}$ must be meticulously optimized to minimize a cost function, denoted by $C$, which quantifies the discrepancy between the training data and the ESN's output.

The cost function and its minimization are formally represented as:

\begin{equation}
    C[\textbf{W}^{(\textrm{out})}]=\frac{1}{n_{\textrm {tr}}} \sum_{t=1}^{n_{\textrm {tr}}} (\textbf{W}^{(\textrm{out})}\bm{r}_t- \bm{c}_t )^2 + \beta \sum_{i=1}^{l} \| W^{(\textrm{out})}_i \|_2^2
    \label{ESN1}
\end{equation}

Here, $W^{(\textrm{out})}_i $ represents the $i$-th row of $\textbf{W}^{(\textrm{out})}$, and $| |_2^2$ signifies the $L_2$ norm. The regression problem is resolved by:
\begin{equation}
\textbf{W}^{(\textrm{out})}\ast = \textbf{Y} \textbf{S}^{\textrm{T}}(\textbf{S}\textbf{S}^{\textrm{T}}+\beta \textbf{I}\textbf{d})^{-1}.
\label{ESN2}
\end{equation}

The ESN's hyperparameters include the reservoir size $N$, the node density $D$, the spectral radius of the reservoir $\rho(\textbf{W}^{(\textrm{r})}\eta)$, the leakage rate $\alpha$, and the Tikhonov regularization parameter $\beta$ in the cost function $C$. Collectively, these are captured in the vector $\bm{\eta}=(N,D,\rho,\alpha,\beta)$. This training methodology is inherently expeditious relative to other forms of RNNs. Following successful training and optimization of hyperparameters, the ESN operates as an autonomous dynamical system. In this mode, a forecasted compressed snapshot $\bm{c}_{t+1}$ can be iteratively used as the succeeding ESN input to predict $\bm{c}_{t+2}$ and continue thereafter. This mode of operation is recognized as the closed-loop scenario.

%**************************************************************
\section{Results and Discussion}
\label{results}

\subsection{Data driver flow model}
\label{dae}
As mentioned earlier, we first trained a DAE network to extract the feature in lower dimensions. For the same, input and output layers have the same dataset i.e. the snapshots from the DNS consist of $u$-velocity component with a dimension of $1501 \times 109 \times 1$. The neural network has multiple convolution layers combined with a max pooling layer while the decoder section has a convolution layer and up-sampling layers, giving a converging-diverging shape. The features or latent modes are obtained at the end of the encoder section and it has a shape of $24 \times 2 \times 34$, thereby giving a dimensionality reduction of 99\%. This could increase further on the expense of reconstruction loss. TABLE \ref{tab:cae} illustrates the various layers and their shape.  After fixing the DAE architecture, various hyperparameters were obtained by using Bayesian optimization. This particular approach is part of AutoML paradigm \cite{he2021automl}, and it assists in achieving an optimized set of parameters in a relatively smaller number of iterations when compared to the grid or random search \cite{turner2021bayesian}. It is worth mentioning that data was scaled between 0-1 before the training. For the entire process, we used 2000 snapshots, 60\% are used for training, 20\% for validation while remaining 20\% for blind testing. The model has around 1.9 Million trainable parameters and the model reaches a loss of $O(10^{-4})$ for both training and validation set in 50 epochs.

%------------------------------------------------------
\begin{table*}[htb]
 \begin{tabular}{l c c l c} 
\hline\hline
\multicolumn{2}{c}{Encoder} & $\quad\quad\quad\quad\quad$ & \multicolumn{2}{c}{Decoder} \\
%\hline                    
Layer & Output size & & Layer & Output size \cr 
\hline
 Encoder input & $1501\times 109 \times 1$ & & Decoder input & $24\times 2 \times 34$ \cr 
 2D Conv-E1 & $1501\times 109 \times 256$ & &2D Conv-D1 & $24\times 2 \times 34$ \cr 
 Max Pool-E1 & $751\times 55 \times 256$ & & 2D Upsamp-D1 & $48\times 4 \times 34$ \cr 
 2D Conv-E2 & $751\times 55 \times 256$ & & 2D Conv-D2 & $48\times 4 \times 32$ \cr 
 Max Pool-E2 & $376\times 28 \times 256$ & & 2D Upsamp-D2 & $96\times 8 \times 32$ \cr 
2D Conv-E3 & $376\times 28 \times 128$ & & 2D Conv-D3 & $96\times 8 \times 64$ \cr 
Max Pool-E3 & $188\times 14 \times 128$ & & 2D Upsamp-D3 & $192\times 16 \times 64$ \cr 
 2D Conv-E4 & $188\times 14 \times 64$ & & 2D Conv-D4 & $192\times 16 \times 128$ \cr 
 Max Pool-E4 & $94\times 7 \times 64$ & & 2D Upsamp-D4 & $384\times 32 \times 128$ \cr 
  2D Conv-E5 & $94\times 7 \times 32$ & & 2D Conv-D5 & $384\times 32 \times 256$ \cr 
 Max Pool-E5& $47\times 4\times 32$ & & 2D Upsamp-D5 & $768\times 64 \times 256$ \cr 
  2D Conv-E6 & $47\times 4 \times 34$ & & 2D Conv-D6 & $768\times 64 \times 256$ \cr 
 Max Pool-E6 & $24\times 2\times 34$ & & 2D Upsamp-D6 & $1536\times 128 \times 256$ \cr 
  &  &  & Output with 2D Conv & $1536\times 128 \times 1$ \cr 
  & &  & Output with Cropping & $1501\times 109 \times 1$ \cr 
\hline\hline
\end{tabular}
\caption{DAE architecture used at first level for feature extraction. The table summarizes the encoder and decoder architectures (Conv = convolution, Max Pool = max pooling, Upsamp = upsampling). Symbol E2 denotes for example encoder hidden layer No. 2.}
\label{tab:cae}
\end{table*}
%------------------------------------------------------

\begin{figure}[ht!]
\centering
\includegraphics[width=12cm, keepaspectratio]{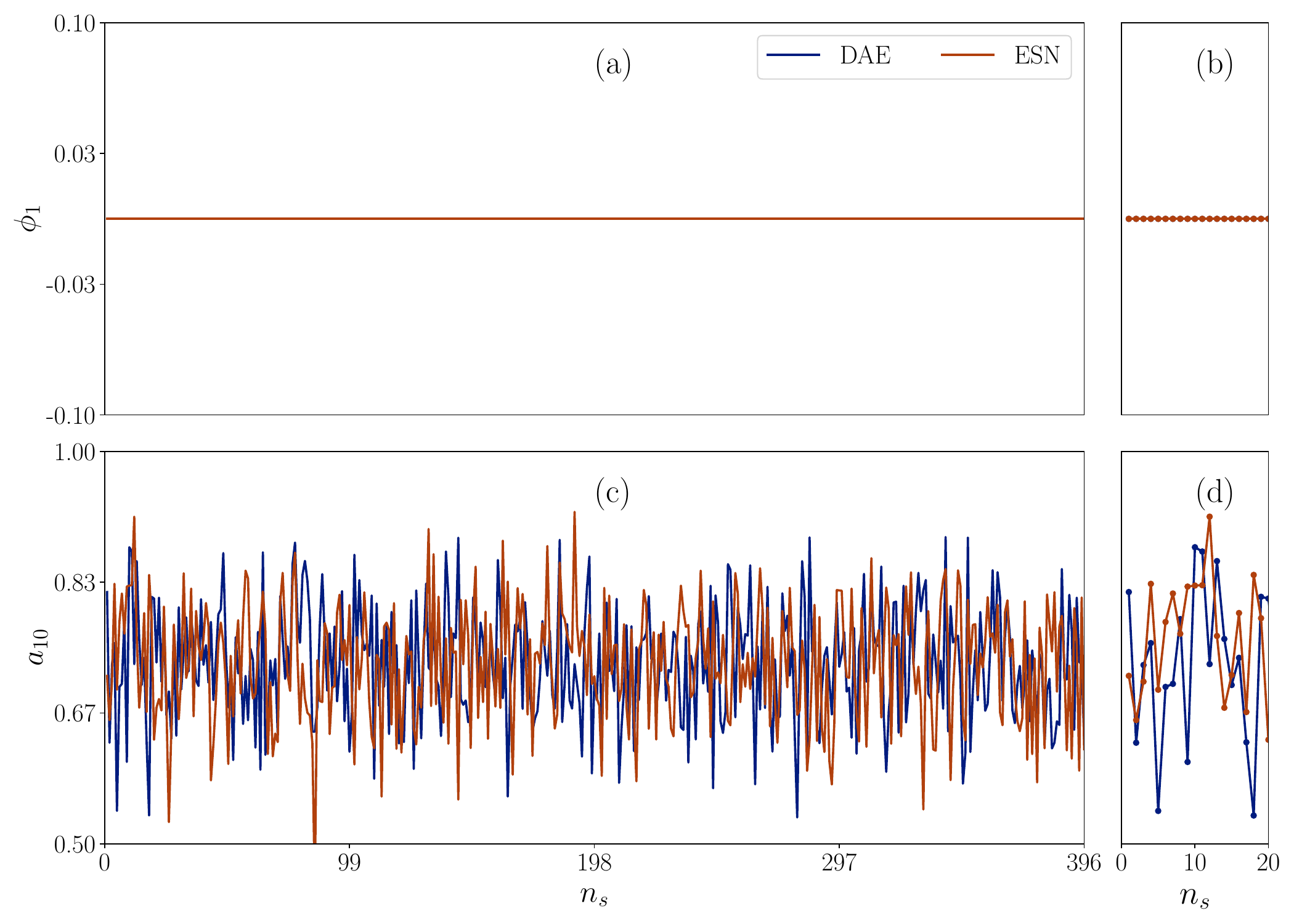}
\caption{\label{auto_results} A comparison of extracted features from DAE which acts as ground truth for the training, and prediction from ESN for 2 of the modes.}
\label{Fig-3}
\end{figure}

\begin{figure}[ht!]
\centering
\includegraphics[width=13cm, keepaspectratio]{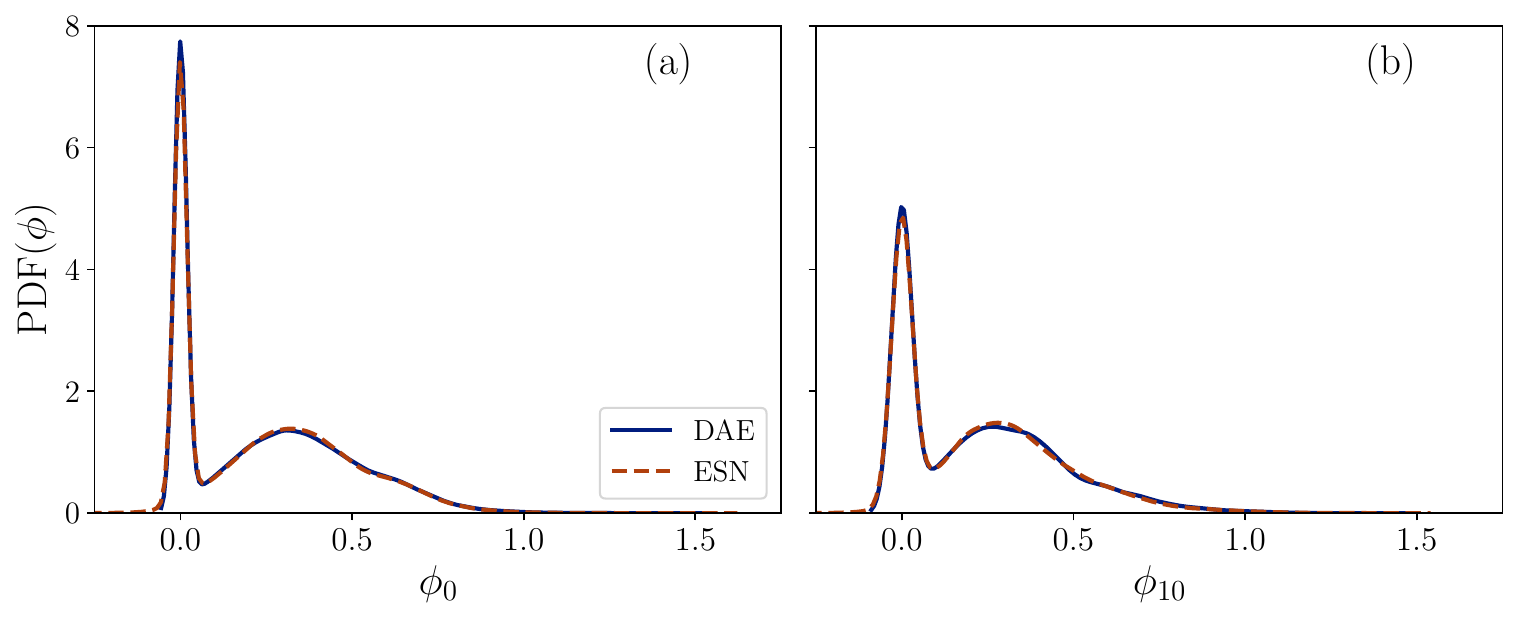}
\caption{\label{auto_results} Comparison between DAE and ESN modes in terms of PDF for 2 of the modes}
\label{Fig-4}
\end{figure}

Once DAE is trained, we extract the features for the entire dataset using its encoder section. These encoded data which have 99\% reduced dimension were further used to train the second level of RNN i.e. ESN. Unlike DAE, training an ESN is relatively easy and computationally cheaper because of straightforward design choices. The optimized network of ESN has 3726 reservoirs, 0.93 as spectral radius, 0.94 as leaking rate and 0.16 as reservoir density. For the best results, the look back history period is 4 timesteps i.e., the model requires 4 timesteps data as an initialization then the system can predict auto-regressively. As a first step of validation, we visualize the low-dimensional features from DAE along with the generated data from ESN on the blind validation set. We call the method generative because the model works auto-repressively once we provide the initial data. Figure \ref{Fig-3} shows such a time series for 2 of the modes out of 1632. It is impressive that the model maintains a value of 0 without any minor fluctuations for $\phi_1$ while the series fluctuates with the right phase and amplitude for $\phi_{10}$. Figure \ref{Fig-4} further illustrates the probability density function (PDF) of these 2 modes for the DAE and ESN. It reaffirms the generative capability of ESN where it is able to match the overall distribution with GT (i.e., DAE for ESN). 

%------------------------------------------------------
%------------------------------------------------------

%------------------------------------------------------
%------------------------------------------------------

\begin{figure}[ht!]
\centering
\includegraphics[width=18cm, keepaspectratio]{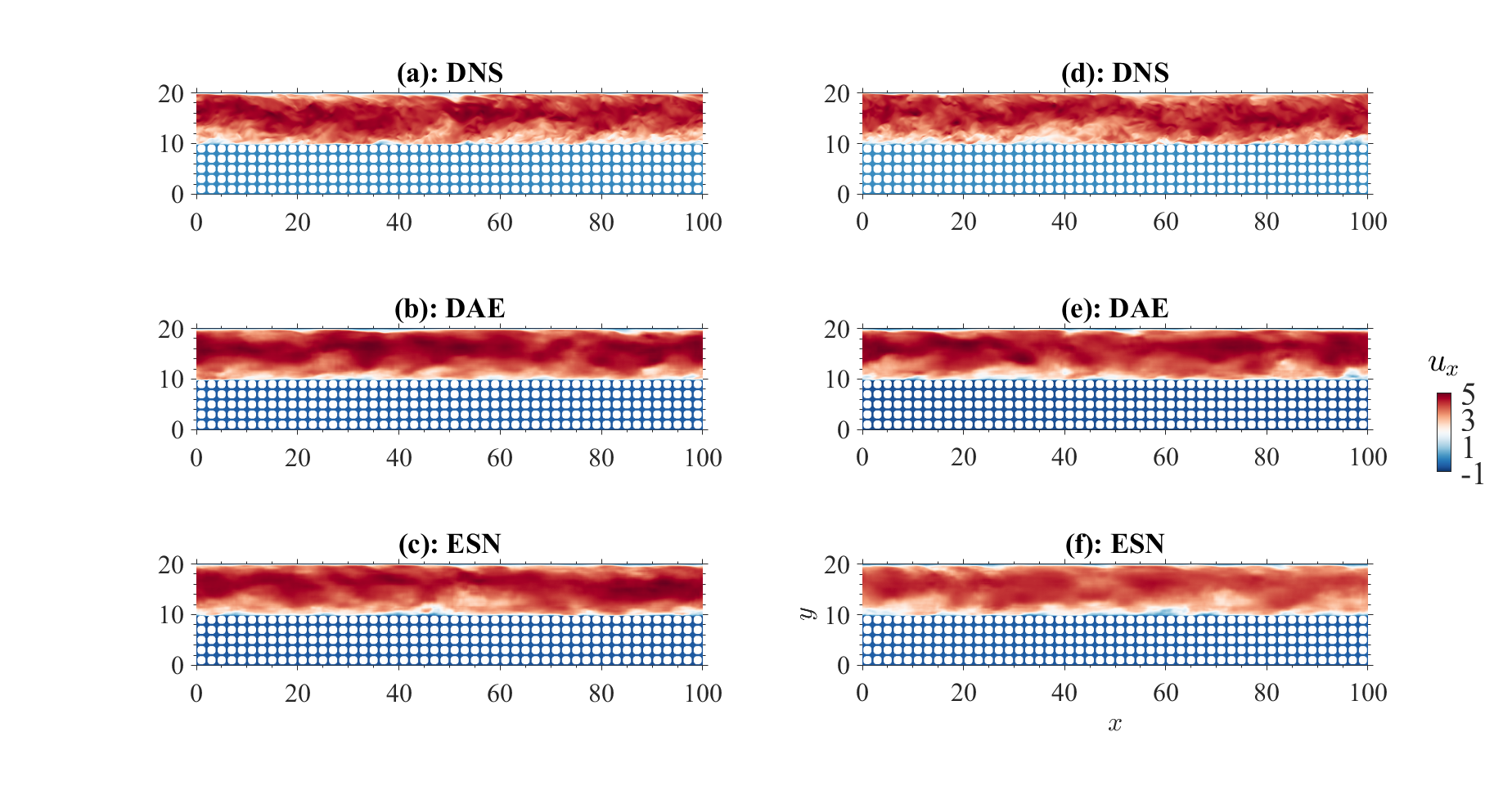}
\caption{\label{auto_results} Contours of $u_x$ by (a,d) DNS (GT), (b,e) DAE (feature extractor) and (c,f) ESN (final end-to-end model) at 2 different timestamps.}
\label{Fig-5}
\end{figure}

\begin{figure}[h!]
\centering
\includegraphics[width=14cm,keepaspectratio]{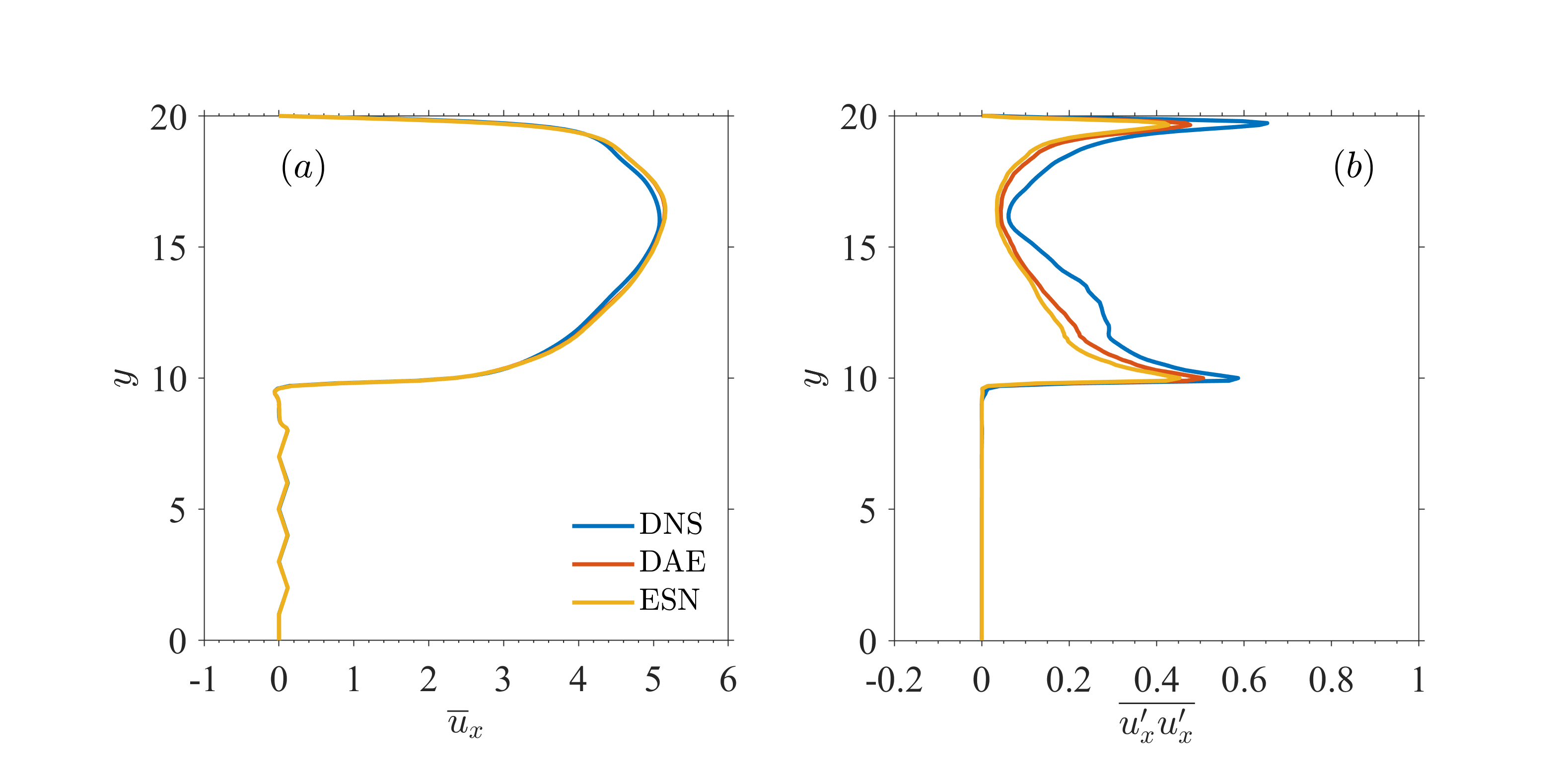}
\caption{\label{auto_results} Comparison among ground truth (DNS), reduced order model (DAE) and temporal model for DAE modes using ESN. (a): Average fields (b) Fluctuations}
\label{Fig-6}
\end{figure}

As a next step of validation, a comparison in high-dimensional space where flow physics needs to be captured is performed. Figure \ref{Fig-5} displays contour plots of the original DNS data in panels (a) and (d), reconstructed data from the features space of DAE in panels (b) and (e), and the decoded value from the prediction of ESN is shown in panels (c) and (f). One can observe a good qualitative agreement in terms of capturing the flow fields. Both regions i.e. main channel flow and porous flow can be predicted. Both DAE and ESN were trained from the DNS, and are not able to fully capture the strong turbulent field in the channel. While the flow in the porous media section is much better because it falls within the laminar region mostly. The obvious reason is the limited capability of DAE to capture the entire flow field because it discards 99\% of input data and uses only 1\% of the features, similar to conventional reduced-order modeling approaches such as POD. The derived model with ESN shows an excellent result while matching the field compared to its GT i.e. DAE. It suggests that the entire model can further be improved if a state-of-the-art feature extractor such as a transformer model can be used instead of DAE. It is worth mentioning that DAE surpasses conventional as well as some advanced feature extractors e.g., variational autoencoder for this use-case, which was observed in our experimentation. Often, averages and fluctuations are critical values for the analysis in industrial applications. Therefore, we further validate the results with mean and fluctuation profiles over the cross-section of the channel. Figure \ref{Fig-6} shows these profiles and the mean velocity profile overlaps with the DNS for both DAE and ESN. The fluctuation of velocity is slightly deviated in the main channel while they remain zero in the porous section due to dominating laminar flow. The distinction made by the model is also great where it can predict the interface where flow is transiting from laminar to turbulent.
%------------------------------------------------------
%------------------------------------------------------

%------------------------------------------------------
%------------------------------------------------------

\begin{figure}[h!]
\centering
\includegraphics[width=13cm,keepaspectratio]{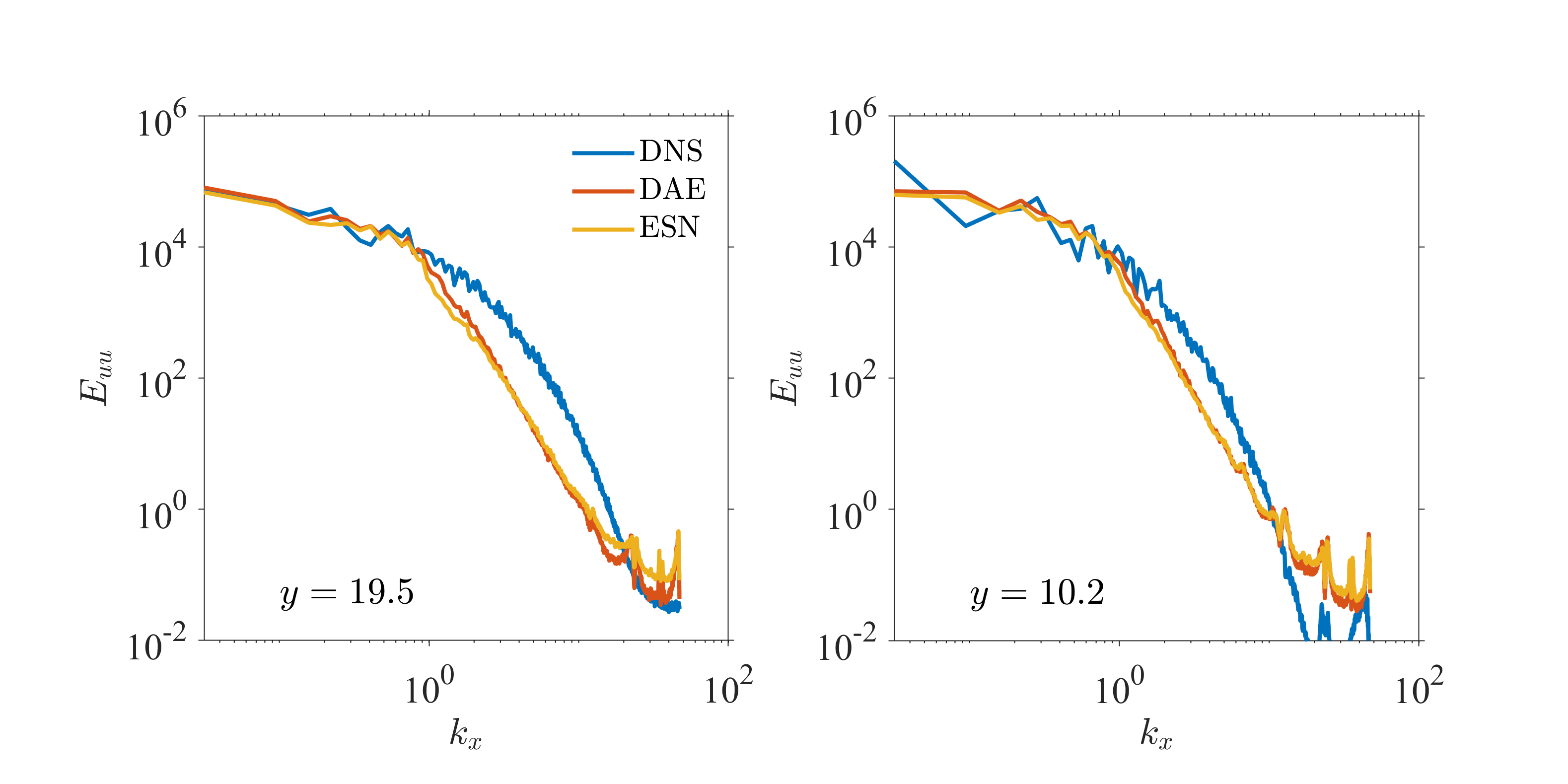}
\caption{Streamwise spectra of velocity fluctuation at different locations, (a) close to the smooth wall, (b) close to the permeable wall}
\label{spectra}
\end{figure}

%To have a thorough comparison, we add results obtained for the Fourier spectra. Figure \ref{spectra} compares the streamwise spectra $E_{uu}$ at the position close to the smooth wall $y=19.5$ as well as close to the permeable wall $y=10.2$. At both locations, DAE and ESN can model and predict the energy at large scale ($k_x < 1$) quite well. However, both DAE and ESN failes predict the rich turbulent kinetic energy in the inertial subrange, especially at $y=19.5$. This is as expected since DAE achieved a dimensionality reduction of 99\%.
For a comprehensive analysis, we include findings from Fourier spectra in our comparison. Figure \ref{spectra} contrasts the streamwise spectra $E_{uu}$ near both the smooth wall at $y=19.5$ and the permeable wall at $y=10.2$. Both the DAE and ESN models perform well in capturing the energy at larger scales ($k_x < 1$). However, neither DAE nor ESN successfully models the abundant turbulent kinetic energy found in the inertial subrange, particularly at $y=19.5$. This limitation is anticipated, given that the DAE model achieves a 99\% dimensionality reduction.

\subsection{Fine tuning to a various porosity}
\label{ft}

The trained model has shown a good agreement in terms of quantitative as well as qualitative characteristics of the flow. However, the trained model is limited to a given data domain that was used for the training which means that the model will not predict accurate results if we use different parametric conditions. This is one of the biggest downsides of such a data-driven model.
Considering the current setup, different porous media with various porosities \citep{Wang2021, Wang.2022} and topologies \citep{Wang2021assessment} are often compared together, whereas the channel remains unchanged.
Variations in porous topology, porosity, and permeability can lead to distinct influences on turbulence modulation, which in turn could impact functionalities like drag reduction, noise absorption, and heat transfer enhancement.
We evaluate the use of fine-tuning as a method to extend the versatility of these models to broader data sets. Fine-tuning involves adapting a pre-trained model to new, limited data. Given that the DAE is computationally intensive within this workflow, we are seeking a substantial reduction of the training time by using fine-tuning. The DAE model was tested using a distinct data set for this assessment.
%There are several approaches which can be employed to increase the applicability of such models to a wider data domain. In this work, we assess the fine-tuning which is widely used. Fine tuning makes the use of existing trained model and tune it with the newer limited data. As mentioned earlier, DAE is a computationally expensive component of this workflow, therefore, we further assess DAE with the fine tuning. We tested the DAE model with a different set of data. 
Here, we used case C06 which has higher level of porosity $\varphi=0.6$. 
The other parameters such as turbulent Reynolds number on both wall sides $Re_\tau^p, Re_\tau^s$, permeability $\sqrt{K_{\alpha\alpha}}^{p+}$ and Forchheimer coefficients $C_{\alpha\alpha}^{{p+}}$ are also quite different between C05 and C06 , as shown in TABLE \ref{tab:1}. 

%\textcolor{red}{I propose to rename cases as follow. I have uploaded case 94, which represent the last one with all frozen layer except one, which you can add here in the figure once you've time :)}

\begin{table*}[htb]
\begin{tabular}{c c c c} 
 \hline
 Case Name & Description & Training time & Mean Square Error \\ [0.5ex] 
 \hline\hline
 DNS & Simulation results which acts as GT for case C06 & - & - \\ 
 \hline

case A & Direct prediction using model from case C05 without any training & 0 & 0.081 \\
 \hline
case B & Model from case C05 with only 1 trainable layer for case C06  & $\mathcal{O}$(0.5 $\times$ 10$^1$) & 0.064 \\
 \hline
case C  & Model from case C05 retrained with 10\% data points for case C06 & $\mathcal{O}$(10$^1$) & 0.049 \\
 \hline
case D  & Model from case C05 retrained with all data points for case C06 & $\mathcal{O}$(10$^2$) & 0.030 \\
 \hline
case E & Model trained from scratch similar to case C05 & $\mathcal{O}$(10$^3$) & 0.025 \\
 \hline
 \hline
\end{tabular}
\caption{Various cases considered for fine-tuning.}
\label{tab:fine_tune}
\end{table*}

TABLE \ref{tab:fine_tune} shows the cases used for the fine-tuning study. It compares the training time and the mean square error of these cases. The case numbering is based on an increasing order of the training time. Case A uses the trained model with C05 without any change. Therefore, its training time is 0 while the error is also the highest. Case E is trained from scratch, which is the same as the procedure in the last section. It is featured with the longest training time and the lowest mean square error showing the high accuracy and high training cost. Cases A and E represent two extreme conditions, whereas cases B, C and D fall between the two extremes. Case D is based on the same model as the original model from C05, however, it is retrained with all data points for case C06. Its training cost $\mathcal{O}$(10$^2$) is one order of magnitude lower than case A whereas the mean square error is only slightly higher which indicates a worthy trade-off. Case C is fundamentally close to case D however only retrained with 10$\%$ data for case C06. Hence, its training time $\mathcal{O}$(10$^1$) is again one order of magnitude lower than case D but with a significant increase of the mean square error. Case B is inherited with the model from case C05 with only 1 trainable layer for case C06. The training time is approximately half of that of case C and shows a significant increase in the error. In general, both case C and case D are valuable approaches with a reasonable trade-off of accuracy and training cost. Case D offers a high accuracy and still a significant reduction of training cost, whereas case C can be valuable when the training resource is extremely limited.

%------------------------------------------------------
\begin{figure}[ht!]
\centering
\includegraphics[width=20cm, keepaspectratio]{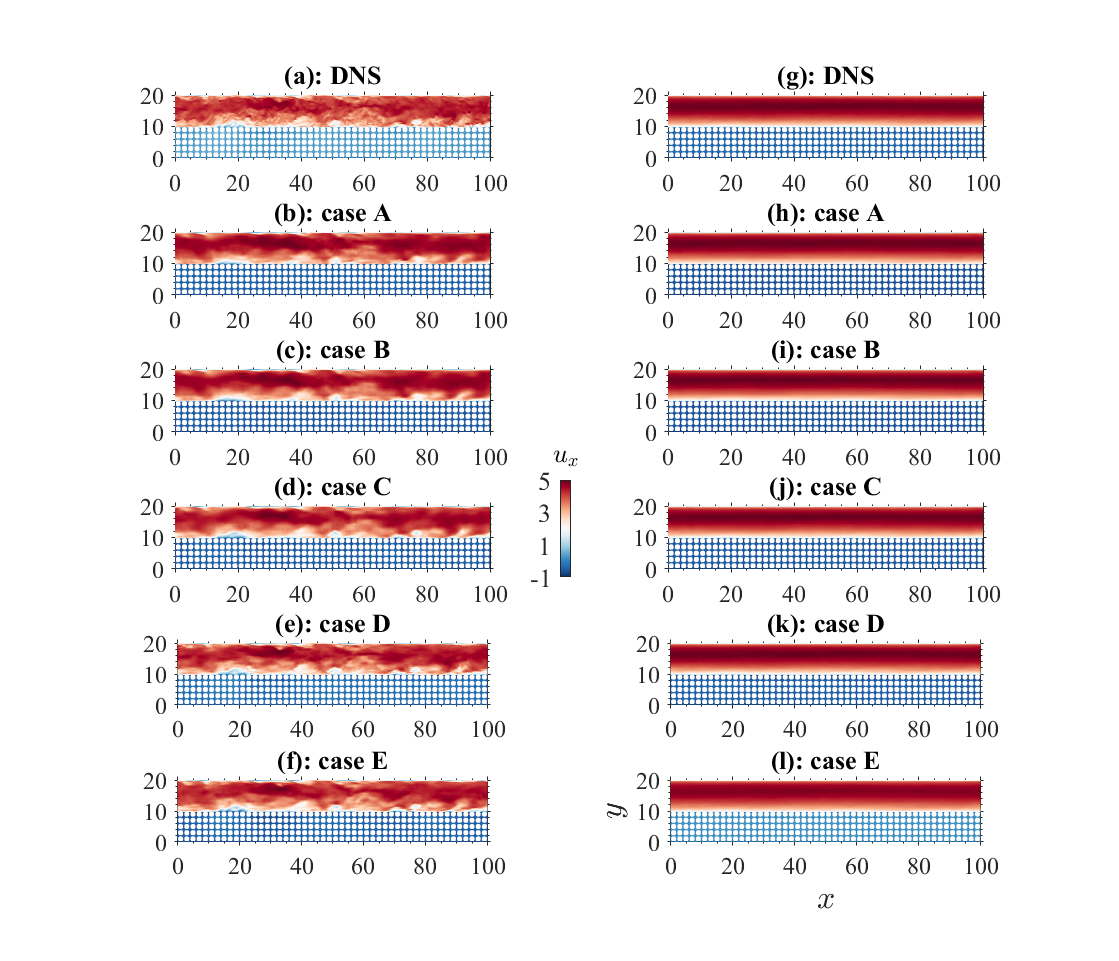}
\caption{\label{auto_results} Contours of (a, b, c, d, e, f): Instantaneous velocity, (g, h, i, j,k,l): temporal average velocity. Labels are as per Tab. \ref{tab:fine_tune}}
\label{Fig-7}
\end{figure} 
%------------------------------------------------------

Figure \ref{Fig-7} compares the instantaneous velocity and time-averaged velocity obtained from the ground truth (DNS) with all the fine-tuning cases from TABLE \ref{tab:fine_tune}. In the DNS result (figure \ref{Fig-7}(a)), turbulent structures on multi-scales can be resolved by fine resolutions. On the other hand, low-rank modeled data by DAE can capture the dominant structures such as the blow and suction events, for instance, the low-velocity region at the interface ($x\approx20, y\approx10$). The fine details are missing due to the truncation of the original data. Compared to the model case A without additional training (figure \ref{Fig-7}(c)), cases B,C,D,E certainly show more details, which corresponds to the accuracy. From cases A to E, an increase in fine motions is observed. On the average results, all the mean velocity contours show similar behavior with the maxima in the channel center and the minima down to the porous domain. It is somehow difficult to tell the quality of modeling.

\begin{figure}[h!]
\begin{subfigure}
\centering
\includegraphics[width=7cm,keepaspectratio]{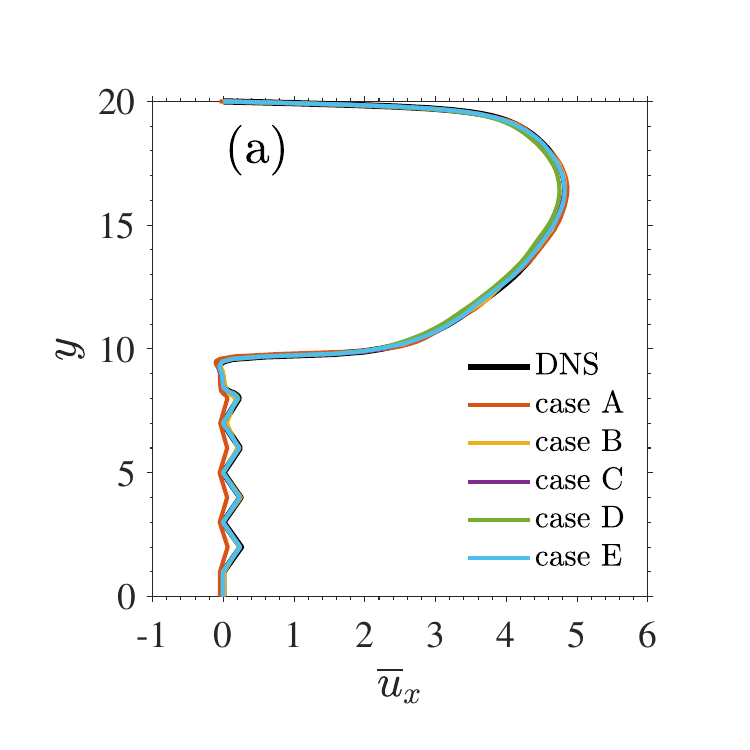}
\end{subfigure}%
\begin{subfigure}
\centering
\includegraphics[width=7cm,keepaspectratio]{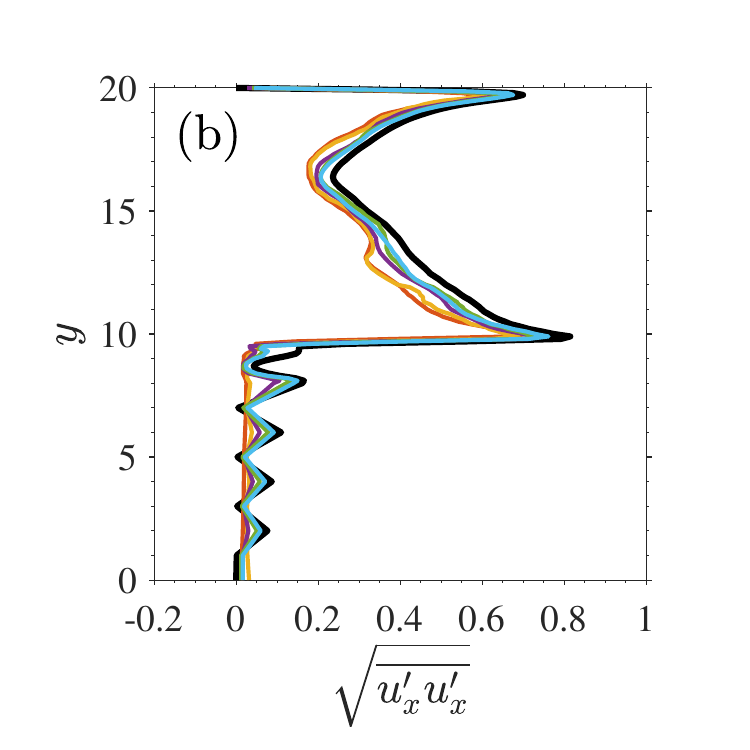}
\end{subfigure}
\caption{\label{auto_results} Comparison among ground truth (DNS), reduced order model (DAE) models. (a): Average fields (b) Fluctuations}
\label{Fig-8}
\end{figure}

Moreover, figure \ref{Fig-8} shows the quantification of the time-averaged streamwise velocity profile and its fluctuation. In the channel region, all the models are able to reproduce the DNS results even in case A without any training on the DNS data. It is because the channel domain is topologically the same even though the porosity in the connected domain is different. A clear difference is seen in the porous domain as well as in the vicinity of the interface. Case A in red color is not able to follow the DNS data of the mean velocity as well as the velocity fluctuation. On the other hand, cases B, C and case D, empowered by the fine-tuning method, show an excellent modeling of the flow inside the porous domain regarding the mean velocity profile. 
This is quite impressive considering case B has only one trainable layer and with more than two orders of magnitude lower training cost.
This suggests that a small amount of data is able to fine-tune the model to model and predict the mean profile in the entire domain.
The prediction of velocity fluctuation in figure \ref{Fig-8}(b) seems much more challenging. Case B shows only a slight improvement but is still close to that of case A. Even though, case B can model the mean velocity profile with no clear discrepancy. 
Case C (10\% new training data and $\mathcal{O}$(10$^2$) reduction of the training time) shows a significant improvement in both the channel and porous region. Its performance is quite close to the case D and E with much higher training cost. Therefore, case C is considered as a balanced case with efficiency and accuracy. On the other hand, a further cost reduction of from case C to case B is not quite significant where the accuracy loss is noticeable.
%It is worthy to mention that all the models including case E can't reproduce the interfacial area quite well which might be a

%A better match is observed in the porous domain. This suggests that a small amount of data is able to fine-tune the model for the porous media domain whereas not quite sufficient for the complex interface- and purely turbulent domain. This illustrates the fact that the multi-scale nature of turbulence is challeging for the fine-tuning rather than the complex topology of the porous media.

%------------------------------------------------------
%------------------------------------------------------
%To further analyze the fine-tuning results in the turbulent channel flow, Figure \ref{spectra2} compares the streamwise spectra $E_{uu}$ of the given fine-tuned cases with DNS. All the fine-tuned have a excellent prediction of the large-scale motion at $k_x<1$. The deviation shows up at large streamwise wavenumber. Even though the difference in the time-averaged velocity fluctuation (Figure \ref{Fig-8}) is observable, it is not that clear to tell in the frequency space with log-log scale. Case C seems to show the best performance, whereas case D is the worst. There is no significant advantage from case A which has actually the lowest MSE error.

\begin{figure}[b!]
	\begin{tabular}{c}
	 \includegraphics[width=0.8\textwidth]{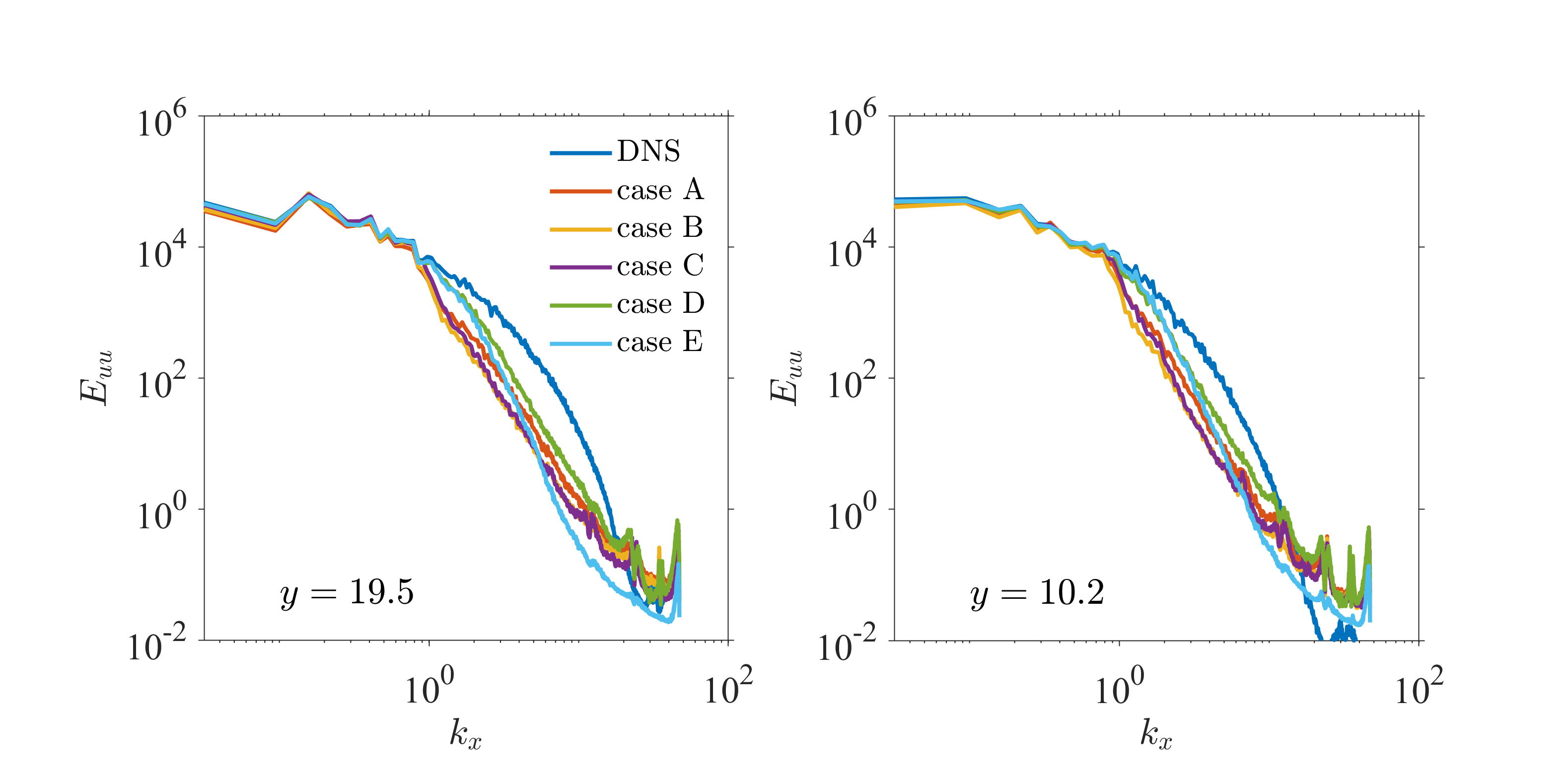}    \\
	\end{tabular}
	\caption{Streamwise spectra $E_{uu}$ close to the smooth wall (a) and permeable wall (b)}
	\label{spectra2}
\end{figure}

For a more in-depth examination of the fine-tuning results in turbulent channel flows, Figure \ref{spectra2} shows the streamwise spectra $E_{uu}$ for all fine-tuned cases against DNS results at $y=19.5$ (close to the top wall) and $y=10.2$ (close to the porous wall). Across all fine-tuned models, the predictions for large-scale motion at $k_x<1$ are highly accurate. However, deviations become apparent at higher streamwise wavenumbers. Although differences in time-averaged velocity fluctuations are noticeable (as seen in figure \ref{Fig-8}), they are less discernible in frequency space when plotted on a log-log scale. Among the cases, cases D and E stand out for their superior performance in the inertial subrange at both positions.
%Among the cases, Case C stands out for its superior performance, while Case D lags behind. Interestingly, Case A, despite having the lowest MSE error, doesn't show any notable advantage.

\begin{figure}[b!]
	\begin{tabular}{c}
	 \includegraphics[width=1\textwidth]{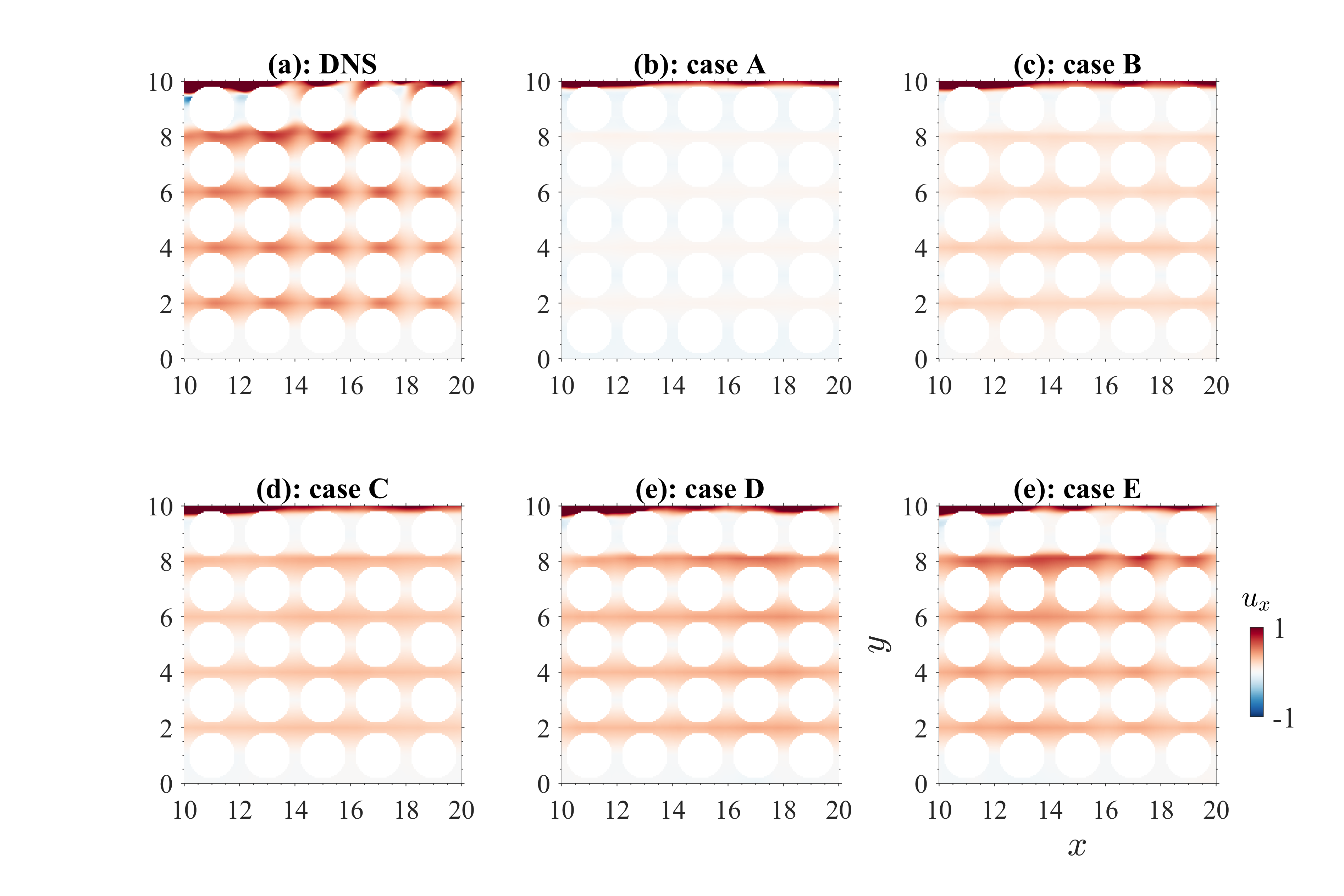}    \\
	\end{tabular}
	\caption{A detailed look of instantaneous velocity $u_x$ to the porous media domain}
	\label{Fig-9}
\end{figure}

%To further analyze the performance of fine-tuned models, the fine-tuned velocity is illustrated in figure\ref{Fig-9}. Case A can capture the flow pattern in the porous domain with the highest accuracy where the fine-tuned case C and case D show also a nice reproduce of the high-speed and low-speed regions inside porous media. On the other hand, case B without any new training has only a blur description of the velocity field inside porous media indicating a poor modeling. Although the flow in porous media is largely effected by the dispersion induced by the porous topology. There are still inhomogeneous velocity appearing at $y\approx8$, which is a laminar-turbulent transitional regime due to its vertical location close to the interface. This has been reproduced by the retrained case A. In addition, the fine-tuned case C also modeled this region fairly well, even though a reduction of training time of one order of magnitude. In region is then only slightly visible in case D. 
To further investigate the performance of fine-tuned models, figure \ref{Fig-9} displays the instantaneous velocity field $u_x$ to the porous media domain. Case A fails to accurately capture the flow pattern within the porous domain due to the missing input of the new training data. 
Starting from case B, fine-tuning is able to significantly increase the modeling of the flow inside porous media even though the training data is quite limited.
Increasing the training effort from case B to case E can progressively enhance the fidelity.
Even though the flow in porous media is significantly influenced by dispersion caused by the porous structure, there are still instances of non-uniform velocity around $y\approx8$. This is indicative of a transitional regime between laminar and turbulent flow, due to its proximity to the interface. This phenomenon is better captured by Cases D and E. 

\begin{figure}[b!]
	\begin{tabular}{c}
	 \includegraphics[width=0.8\textwidth]{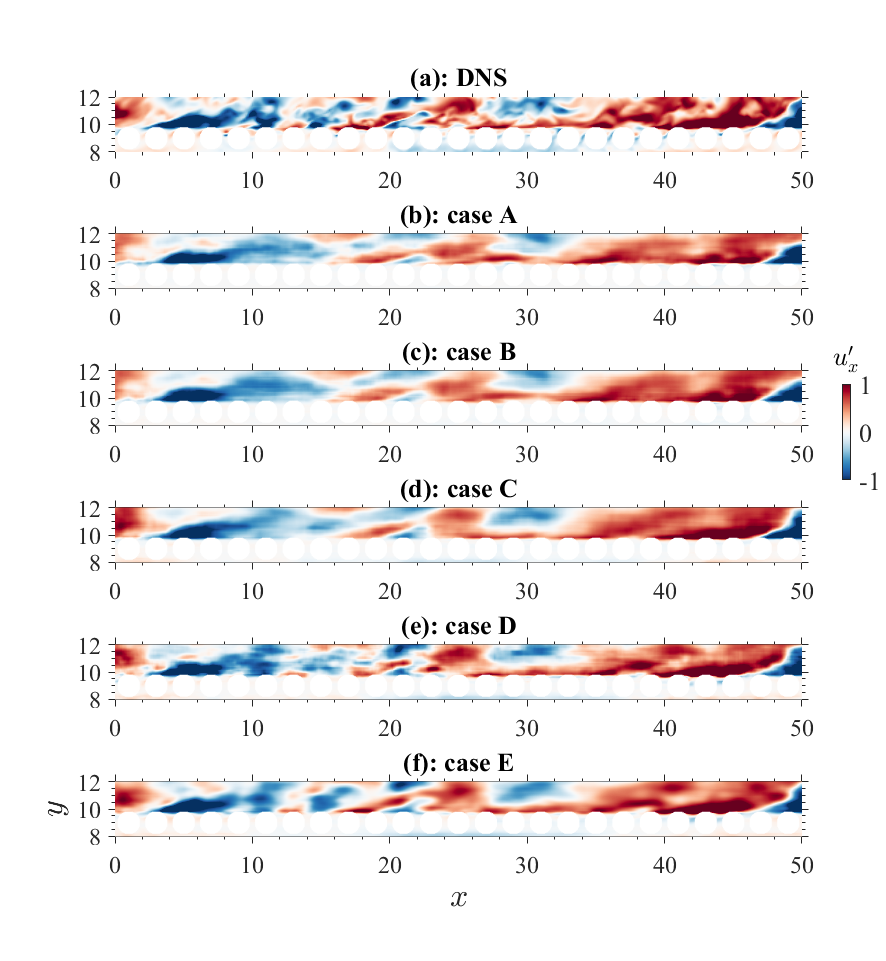}    \\
	\end{tabular}
	\caption{A detailed look of instantaneous velocity fluctuation $u_x^{\prime}$ to the channel flow-porous media interface}
	\label{Fig-10}
\end{figure}

The interfacial behavior is of great interest since it enables bi-directional interactions between channel flow and porous media \citep{Wang2021, Wang.2022}. On the other hand, it is featured with high complexity involving two different types of geometries as well as a mixing state of turbulent, transition and laminar flows. Figure \ref{Fig-10} depicts the instantaneous velocity fluctuation at the interface from DNS and various reduced-order models. In the boundary layer region ($x>10$), DNS shows a rich scale from the large-scale structures to the fine motions, whereas ROM can only reproduce the bigger ones. This is reasonable considering that ROM has to ignore these details to deliver a low-rank model. Among them, case A shows a filtered modeling of the velocity fluctuation field, whereas more details are retained by other cases. Below the interface, the strength of fluctuation is much weaker. But case C, D and E can still deliver the model in a certain way. Case A has received no training data from this DNS, therefore, the porous topology is unclear and the prediction is far from satisfactory.

%**************************************************************
\section{Conclusions and outlook}
\label{concl}

Flows characterized by turbulence over permeable interfaces are ubiquitous and present intricate flow patterns. Addressing the challenges posed by the complex geometry and mixed state of laminar, transitional and turbulent regimes, our study introduces a data-driven, end-to-end machine learning framework specifically designed to model turbulent flows within a channel flow coupled with porous media. To achieve this, we've established a non-linear reduced-order model powered by a deep convolution autoencoder network. Impressively, this model can reduce the spatial dimensions required for highly resolved simulations—integral for direct numerical simulation—by 99\%.
To capture the time-varying characteristics of the reduced modes, we employed a downstream recurrent neural network. This strategic integration allows the model to accurately predict the future trajectories of these modes. Furthermore, to test the adaptability and robustness of our model, we evaluated its performance on a dataset that presents a different porosity from the original training set. Through fine-tuning, our system has demonstrated the potential to significantly cut down on computational resources, reducing training efforts by up to two orders of magnitude when working with a limited dataset (constituting merely 10\% of the full training data). Even with this reduced data input, the results achieved were commendable, particularly with respect to the mean velocity profile.
The efficient fine-tuning of our current model significantly reduces computational time, making it much easier to study the effect of different types of porous media for drag reduction, enhanced heat transfer and noise absorption. This efficiency is particularly advantageous for systematic investigations into interface engineering, allowing for more focused and quicker research cycles.
Of particular note is the observation that the fine-tuned model excels within the porous domain in comparison to the channel and interface regions. This suggests that the inherent topological features of the porous media are more amenable to training compared to the challenging multi-scale attributes typical of turbulent flows.

%**************************************************************
\begin{acknowledgments}
%The work is supported by the Deutsche Forschungsgemeinschaft with Grant No. SCHU 1410/30-1 and in parts by the project “DeepTurb – Deep Learning in and of Turbulence" which is funded by the Carl Zeiss Foundation. 
The work is supported by the funding by Deutsche Forschungsgemeinschaft (DFG, German Research Foundation) project SFB1313 (project No. 327154368) and under Germany’s Excellence Strategy-EXC2075-390740016. In addition, all authors gratefully acknowledge the access to the high performance computing facility Hawk at HLRS, Stuttgart.

\end{acknowledgments}

%**************************************************************
\bibliography{references}

\end{document}